\documentclass[symmetry,review,accept,moreauthors,dvi2pdf,12pt,a4paper]{mdpi}
\lastpage{842}
\doinum{10.3390/sym7020815}
\pubvolume{7}
\pubyear{2015}
\history{Received: 7 March 2015 / Accepted: 19 May 2015 / Published: 29 May 2015}
\pdfoutput=1

\Title{COSMOLOGICAL PROBES FOR SUPERSYMMETRY}

\Author{Maxim Khlopov $^{1,2}$*}

\address{%
$^{1}$ APC laboratory 10, rue Alice Domon et L\'eonie Duquet \\75205
Paris Cedex 13, France\\
$^{2}$ National Research Nuclear University "MEPHI"\\ (Moscow State Engineering Physics Institute), \\
Kashirskoe Sh., 31, Moscow 115409, Russia}

\corres{khlopov@apc.univ-paris7.fr, tel. +33676380567 fax+33 (0)157 279395}
\abstract{The multi-parameter character of supersymmetric dark-matter models implies the combination of their experimental studies with astrophysical and cosmological probes. The physics of the early Universe provides nontrivial effects of non-equilibrium particles and primordial cosmological structures. Primordial black holes (PBHs) are a profound signature of such structures that may arise as a cosmological consequence of supersymmetric (SUSY) models. SUSY-based mechanisms of baryosynthesis can lead to the possibility of antimatter domains in a baryon asymmetric Universe. In the context of cosmoparticle physics, which studies the fundamental relationship of the micro- and macro-worlds, the development of SUSY illustrates the main principles of this approach as the physical basis of the modern cosmology provides cross disciplinary tests in physical and astronomical studies. }

\keyword{Cosmology; Particle physics; Cosmoparticle physics; Supersymmetry; Primordial Black holes; Non-equilibrium particles; Antimatter}

\usepackage[english]{babel}
\usepackage[utf8x]{inputenc}
\usepackage{amsmath}
\usepackage{graphicx}
\usepackage[colorinlistoftodos]{todonotes}

\def\g{{\,\rm g}}
\def\eV{\,{\rm eV}}

\def\({\left(}
\def\){\right)}

\def\beq{\begin{equation}}
\def\eeq{\end{equation}}
\def\bear{\begin{eqnarray}}
\def\ear{\end{eqnarray}}

\begin{document}

\section{Introduction}

The multi-parameter space of Supersymmetric (SUSY) models inevitably implies a combination of direct and indirect studies, in which cosmological probes play an important role.  Such probes  provide supplementary tools to specify the expected properties of SUSY particles both for their efficient search at accelerators and for their ability to explain cosmological dark matter. In the latter case the existence of new conservation laws in SUSY models, such as R-parity, leads to the stability of the lightest particle that possesses the corresponding conserved property. Created in the early Universe such particles should survive to the present day and play the role of dark-matter candidates. This is the simplest and most widely known cosmological impact of Supersymmetry.

However the fundamental basis of supersymmetric models contains a much wider field of cosmological consequences. This basis can naturally include physical mechanisms for inflation and baryosynthesis, and could become the physical grounds for modern inflationary cosmology with baryosynthesis and dark matter. Such mechanisms imply physics of the very early Universe, which can hardly be probed by direct experimental means, hence the necessity to extend the set of theoretical tools to study these phenomena.
Moreover the absence of positive results in the experimental search for SUSY particles at accelerators can move their mass scale to values so high that only indirect methods of study would be appropriate.

Here, following the general framework of discussion of cosmological effects \cite{DMRev,PBHrev}, we consider some nontrivial features  of such effects in their possible application to supersymmetric models, paying special attention to predictions of nonequilibrium particles and primordial nonlinear structures, in which Primordial Black Holes (PBH), effects of PBH evaporation and  even antimatter domains in the baryon asymmetrical Universe can provide nontrivial probes for Supersymmetry in astronomical observations and space experiments.

After a brief review of the cosmological traces of particle models and of their possible relationship with SUSY (Section \ref{Cosmophenomenology}), we turn to PBHs that can play the role of a unique theoretical tool in studies of physics of the very early Universe. We consider some mechanisms of PBH formation (Section \ref{hep}) and possible effects of SUSY particles in PBH evaporation (Section \ref{gravitino}). Metastable SUSY particles or evaporating PBHs are the source of nonequilibrium particles after Big Bang Nucleosynthesis that can influence the primordial chemical composition or provide a non-thermal production of dark  matter particles (Section \ref{evaporation}). A succession of phase transitions in the inflationary Universe can result in  primordial nonlinear structures (Section \ref{MBHwalls}). A profound signature of a nonhomogeneous SUSY-based mechanism for baryosynthesis is the appearance of antimatter domains in a baryon asymmetric Universe (Section \ref{antimatter}). We conclude that the development of supersymmetric models  
 perfectly illustrates the might of the basic principles of cosmoparticle physics \cite{ADS,MKH,book,Khlopov:2004jb,bled,newBook}, which studies the physical basis of the modern cosmology and combines probes from cosmology, astrophysics and experimental physics in a systematic way (Section \ref{Discussion}).

\section{Cosmological traces of particle models}\label{Cosmophenomenology}

Let's start following \cite{DMRev,PBHrev}  (see also \cite{bertone,LSSFW,Gelmini,Aprile:2009zzd,Feng:2010gw}) with a general discussion of a variety of cosmological consequences of particle models and of their possible implementation in the SUSY case.

If R-parity is conserved, the Lightest Supersymmetric Particle (LSP) should be stable. If these particles were created at the early stages of cosmological evolution, they should survive in the Universe until the  present time.
Therefore the simplest cosmological effect of supersymmetric models is a gas of new stable
massive particles, originated from the early Universe.

For particles
with mass $m$, at high temperature \footnote{Throughout the paper we use the units $\hbar=c=k=1$, if it is not specified otherwise .} $T>m$ the equilibrium condition,
$n \cdot \sigma v \cdot t > 1$ is valid, if their annihilation cross
section $\sigma > 1/(m m_{pl})$ is sufficiently large to establish
equilibrium. At $T<m$ such particles go out of equilibrium and their
relative concentration freezes out. Weakly interacting species
decouple from the plasma and the radiation at $T>m$, when $n \cdot \sigma v
\cdot t \sim 1$, i.e. at $T_{dec} \sim (\sigma m_{pl})^{-1}$. This
is the main idea of the calculation of primordial abundances for
WIMP-like dark matter candidates such as the neutralino (see e.g.
\cite{book,newBook} for details). 

The maximal
temperature which is reached in an inflationary Universe is the
reheating temperature, $T_{r}$, after inflation. So, very weakly
interacting particles with annihilation cross section $\sigma <
1/(T_{r} m_{pl})$, as well as very heavy particles with mass $m \gg
T_{r}$ cannot be in thermal equilibrium, and the detailed mechanism
of their production should be considered to calculate their
primordial abundances.  This is the case for the gravitino, which has a semigravitational interaction cross section.

By definition, primordial stable particles survive to the present day
and should be present in the modern Universe. The net effect of
their existence is given by their contribution to the total
cosmological density. If they dominate at the matter dominated stage they play the role of cosmological dark matter that formed the large scale structure of the Universe.
To be a dark matter candidate, stable particles should decouple from the plasma and the radiation before the beginning of matter-dominated (MD) stage. This can be satisfied for a very wide range of cross sections of particle interaction with matter - from superweak to strong. The miracle of
Weakly Interacting Massive particles (such as neutralino) is that their annihilation cross section provides the frozen out concentration that explains the observed dark matter density. For the cross sections typical of weak interaction, a direct search for primordial dark-matter particles in underground experiments is possible (see \cite{DAMA-review,CDMSi,xenon,lux} and references therein).  

Dark-matter particles can annihilate in the Galaxy, contributing by their annihilation products to the flux of cosmic rays and gamma radiation. It was first found in \cite{ZKKC} that such an effect of dark matter annihilation can provide a sensitive tool in indirect studies of dark matter by the measurements of cosmic positron and gamma background fluxes. 

There  may be several types of new stable particles. Then multi-component dark matter scenarios are realized. 
It is interesting that such a multi-component scenario can naturally arise in the heterotic-string phenomenology,  which can embed
both LSP and
other possible types of stable particles or objects.

 The mechanism of compactification and symmetry breaking leads to the
prediction of homotopically stable objects \cite{Kogan1} and
shadow matter \cite{Kogan2}, giving a wide variety of new types of dark matter candidates. 

The embedding of the symmetry of the
Standard model and Supersymmetry within the heterotic string
phenomenology can be also accompanied by  the prediction of a fourth generation of quarks and
leptons \cite{Shibaev} with a stable massive 4th neutrino
\cite{Fargion99} and possibly a stable quark. 

Indeed, the comparison between the rank of the
unifying group $E_{6}$ ($r=6$) and the rank of the Standard model
($r=4$) implies the existence of new conserved charges and new
(possibly strict) gauge symmetries. A new strict gauge U(1) symmetry
(similar to the U(1) symmetry of electrodynamics) is possible, if it
is ascribed to the fermions of the 4th generation. This hypothesis
explains, in particular, the difference between the three known types of neutrinos
and the neutrino of the 4th generation. The latter possesses a new gauge
charge and, being a Dirac particle, cannot have a small Majorana mass
due to see-saw mechanism. If the 4th neutrino is the lightest
particle of the 4th quark-lepton family, the strict conservation of
the new charge makes it absolutely stable.
Following this hypothesis \cite{Shibaev} the quarks and the leptons of the 4th
generation are the source of a new long-range interaction
($y$-electromagnetism), similar to the electromagnetic interaction
of ordinary charged particles. 

New particles with an electric charge and/or a strong interaction can
form anomalous atoms and be contained in the ordinary matter as anomalous
isotopes. For example, if the lightest quark of 4th generation is
stable, it can form stable charged hadrons, serving as nuclei of
anomalous atoms of e.g. anomalous hydrogen and helium
\cite{BKS,BKSR,BKSR1,BKSR2,BKSR3,BKSR4}. The experimental upper limits on the concentration of anomalous isotopes and especially of anomalous hydrogen severely constrains the possibility of new stable charged particles and practically rules out such particles with charges +1 or -1. However, particles with charge -2 can exist and avoid these constraints and be hidden in O-helium atoms - their neutral atom-like bound states with primordial helium nuclei. The O-helium hypothesis can explain some puzzles of dark matter searches, challenging experimental searches for stable doubly charged particles at the LHC \cite{DMRev,DADM}. The -2 charge component of O-helium can originate from stable anti-U quarks of a 4th generation, which can form a $\bar \Delta$-like ($\bar U$$\bar U$$\bar U$) state bound by chromoelectric forces \cite{BKSR3}.

Therefore together with SUSY candidates some other types
of dark matter candidates - massive stable 4th generation neutrinos, as well as nuclear-interacting O-helium dark atoms, made of stable 
(anti-)U quarks of the 4th generation, are possible in the heterotic string phenomenology, the framework of which favors   a multicomponent analysis of dark-matter effects. 

In the multi-component dark matter scenario a detailed study
of the distribution of particles in space, of their condensation in galaxies,
of their capture by stars, Sun and Earth, as well as of the effects of
their interaction with matter and of their annihilation, provides
a sensitive probe for the existence of even
subdominant components. In particular,
though stable neutrinos of the 4th generation with masses about 50
GeV are predicted to be the subdominant form of dark
matter, contributing less than 0,1 \% to the total dark matter density
\cite{ZKKC,Fargion99,DKKM,Grossi,Belotsky,Belotsky2,BKS1,BKS2,BKS3}, this possibility can be ruled out by direct experimental searches for WIMPs  (see
\cite{DAMA-review,CDMSi,xenon,lux} and references therein) and by studies of the effects of the
annihilation of 4th generation neutrinos and antineutrinos in the Galaxy
in the galactic gamma-background. 4th generation neutrino
annihilation inside the Earth should also lead to a flux of underground
monochromatic neutrinos of known types, which can be excluded by the
analysis of the existing data from underground
neutrino detectors.

Primordial unstable particles with a lifetime shorter than the age of
the Universe, $\tau < t_{U}$, cannot survive to the present day.
But, if their lifetime is sufficiently large to satisfy the
condition $\tau \gg (m_{pl}/m) \cdot (1/m)$, their existence in the
early Universe can lead to direct or indirect traces. The cosmological
flux of decay products contributing to the cosmic- and gamma-ray
backgrounds represents the direct trace of unstable particles. If
the decay products do not survive to the present time their
interaction with matter and radiation can cause indirect effects in
the light element abundance or in the fluctuations of thermal
radiation.

If the particle lifetime is much shorter than $1$ s multi-step indirect
traces are possible, provided that particles dominate in the
Universe before their decay. During the dust-like stage of their dominance
black hole formation takes place, and the spectrum of such primordial
black holes traces the particle properties (mass, frozen concentration,
lifetime) \cite{polnarev}. Particle decay in the end of the dust-like
stage influences the baryon asymmetry of the Universe. In any case logical
chains within a given cosmic phenomenology link the predicted properties of even
unstable new particles to the effects accessible in astronomical
observations. Such effects may be important for the analysis of the
observational data.

The parameters of new stable and metastable particles are also
determined by the pattern of symmetry breaking. This pattern
is reflected in a succession of phase transitions in the early
Universe. First order phase transitions proceed through bubble
nucleation, which can result in black hole formation \cite{hawking3} (see e.g.
\cite{kkrs} and \cite{book2} for review and references). Phase
transitions of the second order can lead to the formation of topological
defects, such as walls, strings or monopoles. The observational data
put severe constraints on magnetic monopoles \cite{kz,preskill} and cosmic
wall production \cite{okun}, as well as on the parameters of cosmic
strings \cite{zv1,zv2}. The structure of the cosmological defects can be
changed in successive phase transitions. More complicated forms
such as walls-surrounded-by-strings can appear. Such structures can be
unstable, but their existence can leave a trace in the nonhomogeneous
distribution of dark matter and give rise to large scale structures
of nonhomogeneous dark matter such as {\it archioles}
\cite{Sakharov2,kss,kss2}. Primordial Black Holes, whose hypothetical existence was first formulated by Zeldovich and Novikov \cite{ZN},  represent a
profound signature of such structures.

\section{The reflection of high-energy physics in the PBH spectrum}\label{hep}

Primordial Black Holes (PBHs) are a very sensitive cosmological
probe of physics phenomena occurring in the early Universe. They
could be formed by many different mechanisms (see e.g.\cite{DMRev,PBHrev}
for a review and reference).

After formation PBHs should remain in the Universe and, if they survive to
the present day, they should represent a specific form of dark matter (see e.g. \cite{MPLAPBH}). PBH evaporation by
Hawking radiation \cite{hawking4} makes them a source of
products of evaporation, which contain any type of particles that can exist in our space-time, both known and unknown. For a wide range of parameters the
predicted effect of PBHs contradicts the data and puts
restrictions on the mechanisms of PBH formation and on the underlying
physics of the very early Universe. On the other hand, for some fixed
values of their parameters, PBHs or their evaporation can
provide a nontrivial solution to some astrophysical problems.

Here we outline, following \cite{DMRev,PBHrev,MPLAPBH}, the relationship of new physics with the mechanisms of PBH formation and the possible reflection of its parameters in the PBH spectrum.

\subsection{PBHs from superheavy metastable particles}

The formation of a black hole is highly improbable in a
homogeneous expanding Universe, since it implies metric fluctuations
of order 1. For Gaussian metric fluctuations 
with a dispersion $\left\langle \delta^2 \right\rangle \ll
1$, the probability of such fluctuations is
determined by the exponentially small tail of the high-amplitude part of
the distribution. This probability can even be more suppressed in the
case of non-Gaussian fluctuations \cite{PBHrev}.

If the Universe has an equation of state $p=\gamma \epsilon,$ with the numerical
factor $\gamma$ being in the range $0 \le \gamma \le 1$, the probability to
form a black hole from fluctuations within the cosmological horizon is
given by \cite{carr75}
\begin{equation}
\label{ProbBH}W_{PBH} \propto \exp \left(-\frac{\gamma^2}{2
\left\langle \delta^2 \right\rangle}\right).
\end{equation}
It provides an exponential sensitivity of the PBH spectrum to the softening of the
equation of state in the early Universe ($\gamma \rightarrow 0$) or to the
increase of the ultraviolet part of the spectrum of density fluctuations
($\left\langle \delta^2 \right\rangle \rightarrow 1$). These
phenomena can appear as cosmological consequence of particle theory.

It was first noticed in \cite{khlopov0} that the dominance of superheavy metastable particles with lifetime $\tau \ll
1$~s in the Universe before their decay at $t \le \tau$ can
result in the formation of PBHs, remaining in the Universe after the
particles have decayed and keeping some information on the particle properties
in their spectrum.

After reheating, at $T <
T_0=rm$, particles with mass $m$ and relative abundance
$r=n/n_r$ (where $n$ is the frozen-out concentration of particles and
$n_r$ is the concentration of relativistic species) must dominate in the
Universe before their decay. Dominance of these nonrelativistic
particles at $t>t_0$, where
$t_0=m_{pl}/T_0^2$, corresponds to a
dust-like stage with an equation of state $p=0,$ for which the particle
density fluctuations grow as\begin{equation}
\label{dens}\delta(t)=\frac{\delta \rho}{\rho} \propto t^{2/3}
\end{equation}
 and the development of gravitational instability results in the formation
of gravitationally bound systems, which decouple from the general cosmological
expansion at \begin{equation}
\label{decoup}t \sim t_f \approx t_i
\delta(t_i)^{-3/2},\end{equation}  with $\delta(t_f)\sim 1$ for fluctuations, which enter
the horizon at $t=t_i>t_0$ with amplitude $\delta(t_i)$.

The formation of these systems can result in black-hole formation either
immediately after the system decouples from the expansion or as a result
of the evolution of initially formed nonrelativistic gravitationally
bound systems \cite{Kalashnikov,Kadnikov}.

If the density fluctuation is especially homogeneous and isotropic, it
directly collapses to PBH as soon as the amplitude of fluctuation
grows to 1 and the system decouples from the expansion. The probability
for direct PBH formation in the collapse of such homogeneous and
isotropic configurations gives a minimal estimate of PBH formation at the
dust-like stage.

The mechanism
\cite{PBHrev,khlopov0} is
effective for the formation of PBHs with masses in the interval \beq
\label{Mint}M_0 \le M \le M_{bhmax}.\eeq The minimal mass
corresponds to the mass within the cosmological horizon at time $t
\sim t_0,$ when particles start to dominate in the Universe and it
is equal to
\cite{PBHrev,khlopov0}\beq
\label{MBHmin} M_{0} = \frac{4 \pi}{3} \rho t^3_0 \approx
m_{pl}(\frac{m_{pl}}{r m})^2.\eeq
 The maximal mass is indirectly determined by the condition
 \beq
\label{Mconmax}\tau = t(M_{bhmax}) \delta(M_{bhmax})^{-3/2}\eeq that
fluctuations
 at the considered scale $M_{bhmax}$, entering the horizon at $t(M_{bhmax})$
 with an amplitude $\delta(M_{bhmax})$, can manage to grow to the nonlinear stage,
 decouple and collapse before particles decay at $t=\tau.$
 For a scale-invariant spectrum $\delta(M)=\delta_0$ the maximal mass
 is given by
\cite{PBHrev}\beq \label{MBHmax} M_{bhmax} = m_{pl}
\frac{\tau}{t_{Pl}} \delta_0^{-3/2} =m_{pl}^2 \tau
\delta_0^{-3/2}.\eeq
The direct mechanism of PBH formation can also be effective during a prereheating dust-like post-inflational stage 
of inflaton field oscillations \cite{khlopov1}.
\subsection{Spikes from phase transitions during an inflationary stage}

A scale-dependent spectrum of fluctuations, in which the amplitude of
small scale fluctuations is enhanced, can be another factor
increasing the probability of PBH formation. The simplest functional
form of such a spectrum is represented by a blue spectrum with a power
law dispersion \beq\left\langle \delta^2(M) \right\rangle \propto
M^{-k},\eeq with an amplitude of fluctuation growing at small
$M$ for $k>0$. A realistic account of the existence of other scalar fields
together with the inflaton during the period of inflation can give rise to
spectra with distinctive scales, determined by the parameters of
the considered fields and by their interaction.

In a chaotic inflation scenario, the interaction of a Higgs field $\phi$
with the inflaton $\eta$ can give rise to phase transitions during the
inflationary stage, if this interaction induces a positive mass term
$+\frac {\nu^2}{2} \eta^2 \phi^2$. When in the course of slow
rolling the amplitude of the inflaton decreases below a certain critical
value $\eta_c = m_{\phi}/\nu$ the mass term in the Higgs potential \beq
\label{Higgs} V(\phi, \eta)=-
\frac{m^2_{\phi}}{2}\phi^2+\frac{\lambda_{\phi}}{4}\phi^4 +\frac
{\nu^2}{2}\eta^2 \phi^2 \eeq changes sign and a phase transition takes
place. Such a phase transitions during the inflationary stage lead to the
appearance of characteristic spikes in the spectrum of initial
density perturbations. These spike--like perturbations, at scales
that cross the horizon from 60 to 1 $e$-- folds before the end of
inflation reenter the horizon during the radiation or dust-like era
and could in principle collapse to form primordial black holes. The
possibility of such spikes in a chaotic inflation scenario was first
pointed out in \cite{KofLin} and implemented in \cite{Sakharov0} as a
mechanism of PBH formation.

If a phase transition takes place at $e$--folding $N$ before the end
of inflation, the spike re-enters the horizon at the radiation dominance
(RD) stage and forms Black hole of mass \beq \label{Mrd} M \approx
\frac{m^2_{Pl}}{H_0} \exp\{2 N\}, \eeq where $H_0$ is the Hubble
constant during the period of inflation.

If the spike re-enters the horizon during the matter dominance (MD) stage it
should form black holes of mass \beq \label{Mmd} M \approx
\frac{m^2_{Pl}}{H_0} \exp\{3 N\}. \eeq

\subsection{PBHs from a first-order
phase transitions in the early
Universe}\label{phasetransitions}

First-order phase transitions go
through bubble nucleation. The simplest way to describe first order phase transitions
with bubble creation in the early Universe is based on a scalar field
theory with two non degenerate vacuum states. Being stable at the
classical level, the false vacuum state decays due to quantum
effects, leading to a nucleation of bubbles of true vacuum and their
subsequent expansion \cite{PBHrev}. The potential energy of the false
vacuum is converted into the kinetic energy of the bubble walls thus
making them highly relativistic in a short time. The bubble expands
till it collides with another one. As it was shown in
\cite{kkrs} a black hole may be created in a collision of
two bubbles.

Just after the
collision, the mutual penetration of the walls up to a distance
comparable with their width is accompanied by a significant increase in potential
energy \cite{PBHrev}. Then the walls reflect each other and accelerate
backwards. The space between them gets filled by the field in the
false vacuum state, converting the kinetic energy of the wall back to
the energy of the false vacuum state and slowing down the velocity of
the walls. Meanwhile the outer area of the false vacuum is absorbed
by the outer wall, which expands and accelerates outwards.
There is an instant when the central region of the false
vacuum is causally separated fromm the walls \cite{oscilon} . If this false vacuum bag shrinks in its oscillations
within its gravitational radius, a black hole is formed. The mass of this PBH is given
by (see \cite{kkrs})
\begin{equation}
\label{15}M_{BH}=\gamma _1M_{bub}
\end{equation}
where $\gamma _1\simeq 10^{-2}$ and $M_{bub}$ is the mass that could
be contained in the bubble volume at the epoch of collision for a full thermalization of bubbles.

If inflation ends with a first-order phase transition, collisions between bubbles of Hubble size in the percolation regime lead
to copious PBH formation with masses

\begin{equation}
\label{16}M_0=\gamma _1M_{end}^{hor}= \frac{\gamma
_1}2\frac{m_{pl}^2}{H_{end}},
\end{equation}
where $M_{end}^{hor}$ is the mass within the Hubble horizon at the end of
inflation. According to (\cite{kkrs}) the initial mass fraction of these
PBHs is given by $\beta _0\approx\gamma _1/e\approx 6\cdot 10^{-
3}$. For example, for typical value of $H_{end}\approx 4\cdot
10^{-6}m_{pl}$ the initial mass fraction $\beta $ corresponds to
PBHs with masses $M_0\approx 1\g$.

\section{Gravitino production by PBH evaporation}\label{gravitino}

Presently there is no observational evidence proving the existence of
PBHs. However, even the absence of PBHs provides a very sensitive
theoretical tool to study the physics of the early Universe. PBHs represent a
nonrelativistic form of matter and their density decreases with the
scale factor $a$ as $a^{-3} \propto T^{3}$, while the total
density is $\propto a^{-4}\propto T^{4}$ in the period of radiation
dominance (RD). As it must be formed within the horizon, a PBH of mass $M$ can be
formed only after
\begin{equation}\label{tfRD}t(M)=\frac{M}{m_{pl}}{t_{pl}}=\frac{M}{m_{pl}^2}.\end{equation} If they are formed
during the RD stage, the smaller are the masses of PBHs, the larger
their relative contribution to the total density during the modern MD
stage. Therefore, even the modest constraint on the density of PBHs of mass $M$ \beq
\label{OmPBH}\Omega_{PBH}(M)=\frac{\rho_{PBH}(M)}{\rho_{c}}\eeq in
units of critical density $\rho_{c}=3 H^2/(8 \pi G)$ from the
condition that their contribution $\alpha(M)$ to the total
density
\begin{equation}\label{defalpha}\alpha(M)\equiv\frac{\rho_{PBH}(M)}{\rho_{tot}}=\Omega_{PBH}(M)
\end{equation} for $\rho_{tot}=\rho_{c}$ does not exceed that
of dark
matter\begin{equation}\label{DMalpha}\alpha(M)=\Omega_{PBH}(M) \le
\Omega_{DM}=0.23\end{equation} converts into a severe constraint on
their contribution
\begin{equation}\label{defbeta}\beta \equiv
\frac{\rho_{PBH}(M,t_f)}{\rho_{tot}(t_f)}\end{equation} during the
period $t_f$ of their formation. If formed during the RD stage at
$t_f=t(M)$, given by (\ref{tfRD}), which corresponds to the
temperature $T_f=m_{pl}\sqrt{m_{pl}/M}$, PBHs contribute to the
total density at the end of the RD stage at $t_{eq}$, corresponding to
$T_{eq}\approx 1 \eV$,  by a factor
$a(t_{eq})/a(t_f)=T_f/T_{eq}=m_{pl}/T_{eq}\sqrt{m_{pl}/M}$ larger
than in the period of their formation. The constraint on $\beta(M)$,
following from Eq.(\ref{DMalpha}) is then given
by\begin{equation}\label{DMbeta}\beta(M)=\alpha(M)\frac{T_{eq}}{m_{pl}}\sqrt{\frac{M}{m_{pl}}}
\le 0.23 \frac{T_{eq}}{m_{pl}}\sqrt{\frac{M}{m_{pl}}}.\end{equation}

The possibility of PBH evaporation, revealed by S. Hawking
\cite{hawking4}, strongly influences the effects of PBHs. In the strong
gravitational field near gravitational radius $r_g$ of the PBH, the quantum
effect of creation of particles with momentum $p \sim 1/r_g$ is
possible. Due to this effect, the PBH becomes a black-body source of
particles with temperature (in the units
$\hbar=c=k=1$)\begin{equation}\label{TPBHev}T=\frac{1}{8\pi G
M}\approx10^{13} {\rm GeV} \frac{1 {\rm g}}{M}.\end{equation} The
evaporation time scale can be written in the following form
\begin{equation}
\label{evop} \tau_{BH}=\frac{M^3}{g_*m_{pl}^4}
\end{equation}
where $g_*$ is the number of effective massless degrees of freedom. For $M \le
10^{14}$~g, it is less, than the age of the Universe. Such PBHs can not
survive to the present day and the magnitude of Eq.(\ref{DMalpha}) should be re-defined as the contribution to the
total density at the moment of PBH evaporation. For PBHs formed during the
RD stage and evaporated during the RD stage at $t<t_{eq}$, the relationship
Eq.(\ref{DMbeta}) between $\beta(M)$ and $\alpha(M)$ is given by
\cite{polnarev,NovikovPBH}
\begin{equation}\label{DMbetaRD}\beta(M)=\alpha(M)\frac{m_{pl}}{M}.\end{equation}
The relationship between $\beta(M)$ and $\alpha(M)$ has a more
complicated form if PBHs are formed during early dust-like stages
\cite{book,polnarev,polnarev1,khlopov6}, or if such stages take place
after the PBH formation\cite{book,khlopov6}. The relative contribution of
PBHs to the total density does not grow during the dust-like stage and the
relationship between $\beta(M)$ and $\alpha(M)$ is model-dependent. A minimal model-independent factor
$\alpha(M)/\beta(M)$ follows from the enhancement
taking place only during the RD stage between the first second of
expansion and the end of RD stage at $t_{eq}$, since radiation
dominance in this period is supported by the light
element abundances and by the spectrum of the Cosmic Microwave Background (CMB)
\cite{book,newBook,polnarev,polnarev1,khlopov6}

The effects of PBH evaporation make astrophysical data much more
sensitive to the existence of PBHs.
 Constraining the abundance of primordial
black holes can lead to invaluable information on cosmological
processes, particularly as they are probably the only viable probe
of the power spectrum at very small scales, which remain far from
the CMB and from the Large Scale
Structures (LSS) sensitivity ranges. To date, only PBHs with initial
masses between $\sim 10^9$~g and $\sim 10^{16}$~g have led to
stringent limits (see {\it e.g.}
\cite{polnarev,carr1,carrMG,LGreen,Cline:1998fx}) from the consideration of the
entropy per baryon, the deuterium destruction, the $^4$He
destruction and the cosmic rays currently emitted by the Hawking
process \cite{hawking4}. The existence of light PBHs should lead to
important observable constraints, either through the direct effects
of the evaporated particles (for initial masses between $10^{14}$~g
and $10^{16}$~g) or through the indirect effects of their
interaction with matter and radiation in the early Universe (for PBH
masses between $10^{9}$~g and $10^{14}$~g). 

Several constraints on the density of PBHs have been derived in
different mass ranges assuming the evaporation to standard
model particles only: for $10^9~{\rm g}<M<10^{13}~{\rm g}$ the entropy
per baryon at nucleosynthesis  was used \cite{mujana} to obtain
$\beta < (10^9~{\rm g}/M)$, for $10^9~{\rm g}<M<10^{11}~{\rm g}$ the
production of $n\bar{n}$ pairs at nucleosynthesis was used
\cite{153} to obtain $\beta  < 3\times 10^{-17} (10^9~{\rm
g}/M)^{1/2}$ , for $10^{10}~{\rm g}<M<10^{11}~{\rm g}$ deuterium
destruction was used \cite{152} to obtain $\beta  < 3\times 10^{-22}
(M/10^{10}~{\rm g})^{1/2}$, for $10^{11}~{\rm g}<M<10^{13}~{\rm g}$
spallation of $^4$He was used \cite{khlopov6,vainer} to obtain
$\beta  < 3\times 10^{-21} (M/10^9~{\rm g})^{5/2}$, for $M\approx
5\times 10^{14}~{\rm g}$ the gamma and cosmic rays were used
\cite{155,barrau} to obtain $\beta < 10^{-28}$. Slightly more
stringent limits were obtained in \cite{kohri}, leading to $\beta <
10^{-20}$ for masses between $10^{9}~{\rm g}$ and $10^{10}~{\rm g}$
and in \cite{barraugamma}, leading to $\beta < 10^{-28}$ for
$M=5\times 10^{11}~{\rm g}$. Gamma rays and antiprotons were also
recently re-analyzed in \cite{barraupbar} and
\cite{Custodio:2002jv}, improving a little the previous estimates.
Such constraints, related to phenomena occurring after 
nucleosynthesis, apply only for black holes with initial masses
above $\sim 10^9$~g. Below this value, the only limits for a long time was the very
weak entropy constraint (related with the photon-to-baryon ratio).

However, since the evaporation products are created by
the gravitational field, any quantum with a mass lower than the
black hole temperature should be emitted, independently of the
strength of its interaction. This could provide a copious production
of superweakly interacting particles that cannot be in
equilibrium with the hot plasma of the very early Universe. It makes
evaporating PBHs a unique source of all the species which can exist
in the Universe.

To derive a limit in the initial mass range $m_{pl}<M<10^{11}$~g,
gravitinos emitted by black holes\footnote{The limits on the abundance of PBHs from gravitino production were first calculated in \cite{lemoine}.} were considered in \cite{KBgrain}.
Gravitinos are expected to be present in all local supersymmetric
models, which are regarded as the more natural extensions of the
standard model of high energy physics (see, {\it e.g.}, \cite{olive}
for an introductory review).

Following \cite{book,newBook,khlopov6,khlopov7} and
\cite{lemoine,green1} (but in a different framework and using more
stringent constraints),
 limits on the mass fraction of black holes
at the time of their formation ($\beta \equiv
\rho_{PBH}/\rho_{tot}$) were derived in \cite{KBgrain} using the
production of gravitinos during the evaporation process. Depending
on whether gravitinos are expected to be stable or metastable, 
limits are obtained using the requirement that they do not overclose
the Universe or that the formation of light nuclei by the
interactions of $^4$He nuclei with a nonequilibrium flux of D,T,$^3$He 
and $^4$He does not contradict the observations. This approach is
more constraining than the usual study of photo-dissociation induced
by photons-photinos pairs emitted by decaying gravitinos. It opened
a new window for the upper limits on $\beta$ below $10^9$~g. 

\section{Nonequilibrium particles}\label{evaporation}

Superweakly interacting gravitinos could not be in thermal equilibrium and thus the mechanisms of their production gives rise to appearance of nonequilibrium particles during the radiation dominated stage. Metastable particles, decaying via hadronic or electromagnetic channels after Big Bang nucleosynthesis, result in fluxes of photons and electron-positron pairs with energies much larger than the thermal energies in this period.  Nucleon-antinucleon pairs from such decays represent a profound example of the creation of species that cannot be in equilibrium in the considered period. The interaction of primordial nuclei with nonequilibrium particles from electromagnetic and hadronic cascades leads to recoil nuclei and nuclear fragments with energies above the Coulomb barrier, and makes successive nuclear reactions possible.

\begin{figure}
\centering
\includegraphics[scale=0.7]{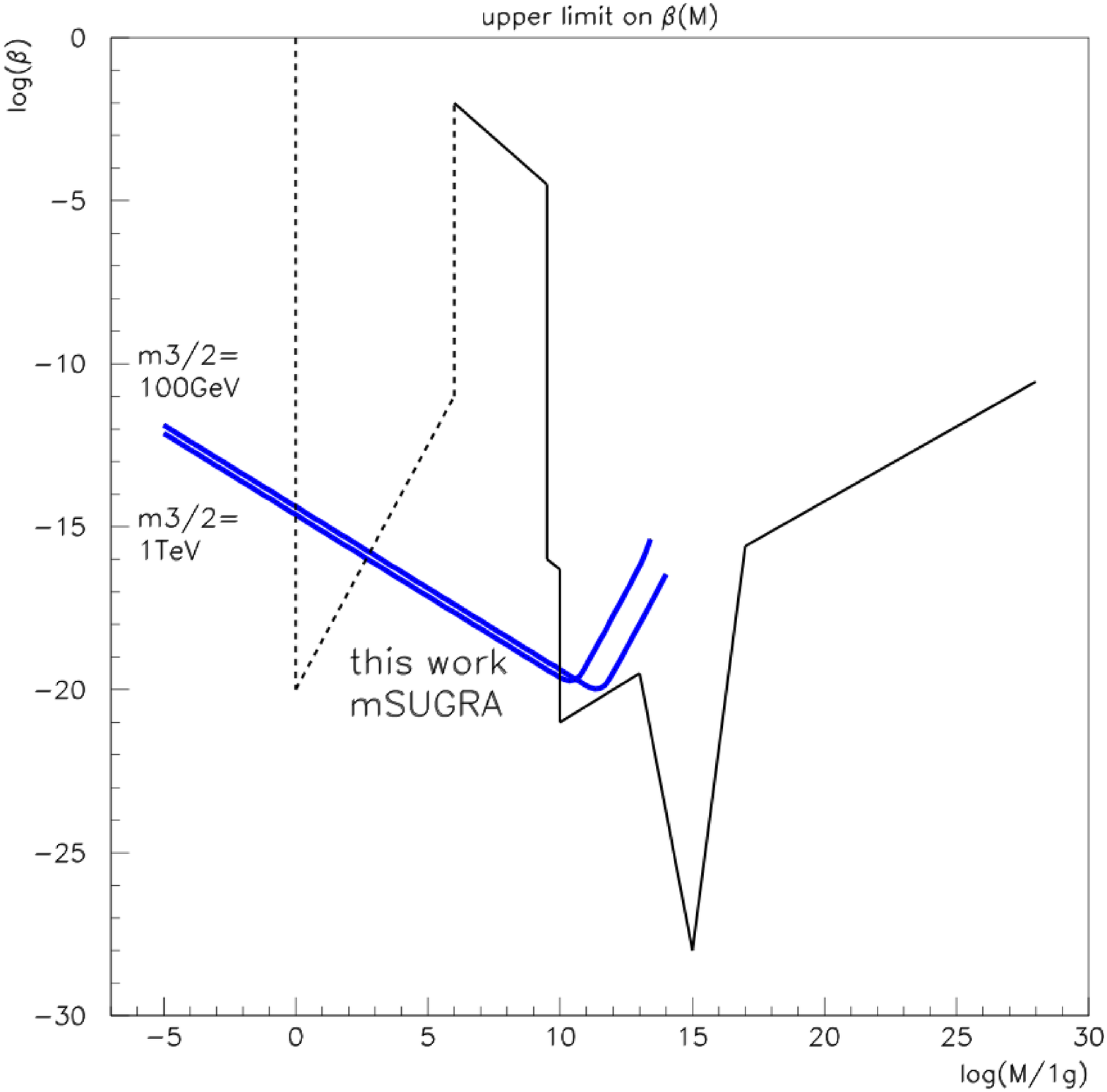}
\caption{\label{fig:beta_NB}Constraints of \cite{KBgrain} on the fraction of the Universe going into PBHs (adapted from
\cite{polnarev,carr1,carrMG,LGreen}). The two curves obtained with
gravitino emission in mSUGRA correspond to $m_{3/2}$ = 100 GeV
(lower curve in the high mass range) and  $m_{3/2}$ = 1 TeV (upper
curve in the high mass range)}
\end{figure}

 In the framework of minimal
Supergravity (mSUGRA), the gravitino mass is, by construction,
expected to lie around the electroweak scale, {\it i.e.} in the 100
GeV range. In this case, the gravitino is {\it metastable} and
decays after nucleosynthesis, leading to important modifications of
the nucleosynthesis paradigm. Instead of using the usual
photon-photino decay channel, the study of \cite{KBgrain} relied on
the more sensitive gluon-gluino channel. Based on
\cite{khlopovlinde,khlopovlinde2,khlopovlinde3,khlopov3,khlopov31},
the antiprotons produced by the fragmentation of gluons emitted by
decaying gravitinos were considered as a source of nonequilibrium
light nuclei resulting from collisions of those antiprotons on
equilibrium nuclei. Then, $^6$Li, $^7$Li and $^7$Be nuclei
production by the interactions of the nonequilibrium nuclear flux
with $^4$He nuclei in equilibrium was taken into account and compared
with data (this approach is supported by several recent analyses
\cite{Karsten,Kawasaki} which lead to similar results). The
resulting Monte-Carlo estimates \cite{khlopov3} lead to the
following constraint on the concentration of gravitinos: $n_{3/2}<
1.1\times 10^{-13}m_{3/2}^{-1/4}$, where $m_{3/2}$ is the gravitino
mass in GeV. This constraint has been successfully used to derive an
upper limit on the reheating temperature 
\cite{khlopov3}: $T_R < 3.8\times 10^6$~GeV. The consequences of
this limit on cosmic-rays emitted by PBHs was considered, {\it
e.g.}, in \cite{barrauprd}. In the approach of \cite{KBgrain} this
stringent constraint on the gravitino abundance was related to the
density of PBHs through direct gravitino emission. The usual
Hawking formula \cite{hawking4} was used for the number of particles
of type $i$ emitted per unit of time $t$ and per unit of energy $Q$.
Introducing the  temperature defined by Eq. (\ref{TPBHev}) $
T=hc^3/(16\pi^2 k G M)\approx(10^{13}{\rm g})/{M}~{\rm GeV}, $
taking the relativistic approximation for $\Gamma_s$, and
integrating over time and energy, the total number of quanta of type
$i$ can be estimated as:
\begin{equation}\label{Niev}N_i^{TOT}=\frac{27\times 10^{24}}{64\pi ^3
\alpha_{SUGRA}}\int_{T_i}^{T_{Pl}}\frac{dT}{T^3}\int_{m/T}^x\frac{x^2dx}{e^x-(-1)^s}
\end{equation} where $T$ is in GeV, $m_{pl}\approx 10^{-5}$~g,
$x\equiv Q/T$, $m$ is the particle mass and $\alpha_{SUGRA}$
accounts for the number of degrees of freedom through
$M^2dM=-\alpha_{SUGRA}dt$ where $M$ is the black hole mass. Once the
PBH temperature is higher than the gravitino mass, gravitinos will
be emitted with a weight related to their number of degrees of
freedom. Computing the number of emitted gravitinos as a function of
the PBH initial mass and matching it with the limit on the gravitino
density imposed by nonequilibrium nucleosynthesis of light elements
leads to an upper limit on the PBH number density. If PBHs are
formed during a radiation dominated stage, this limit can easily be
converted into an upper limit on $\beta$ by evaluating the energy
density of the radiation at the formation epoch. The resulting limit
is shown on Fig. \ref{fig:beta_NB} and leads to an important improvement
over previous limits, nearly independently of the gravitino mass in
the interesting range. This opens a new window on the very small
scales in the early Universe.

\begin{figure}
    \begin{center}
        \includegraphics[scale=0.7]{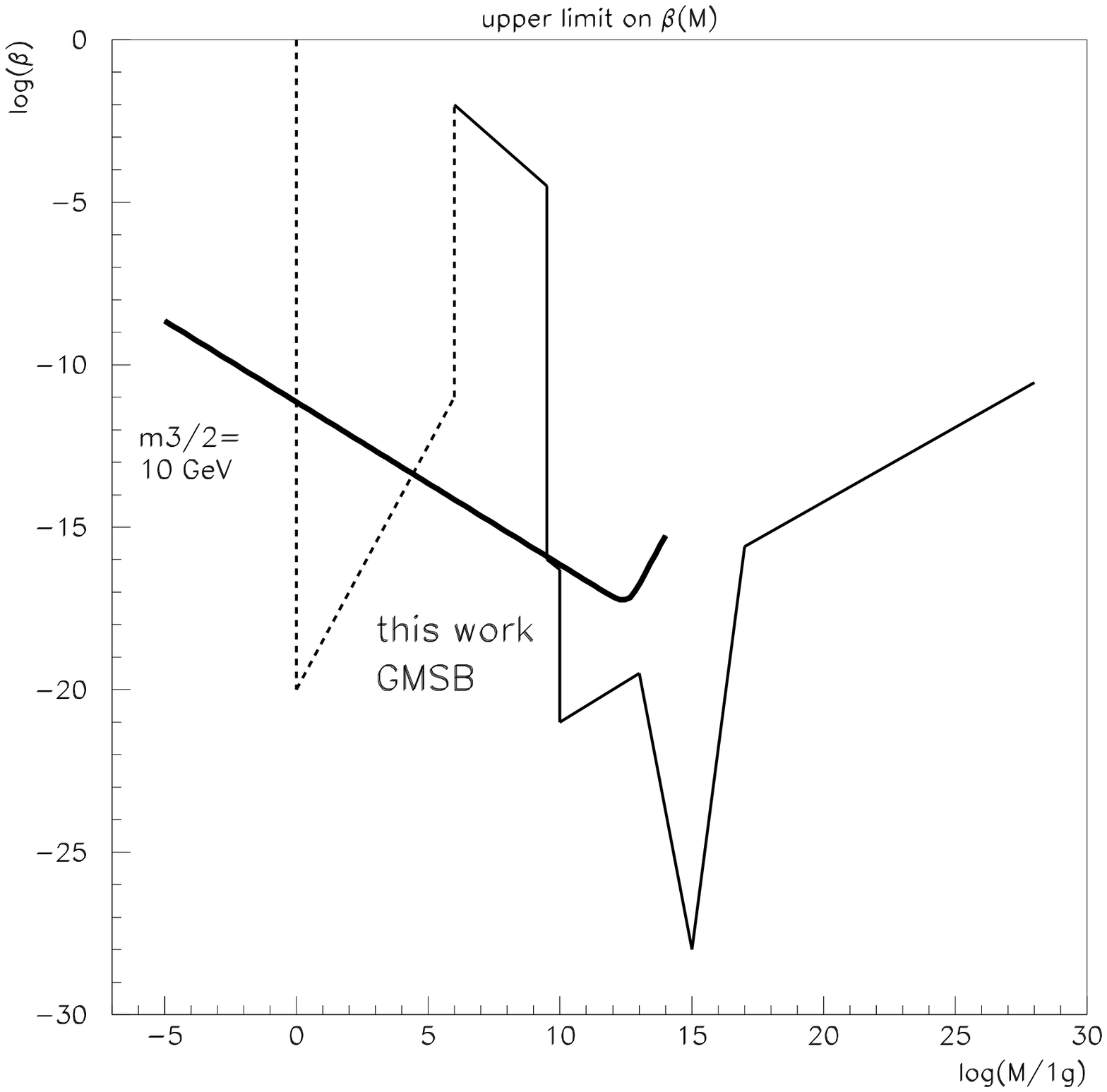}
        \caption{Constraints of \cite{KBgrain} on the fraction of the Universe going into PBHs (adapted from
\cite{polnarev,carr1,carrMG,LGreen}). The curve obtained with
gravitinos emission in GMSB correspond to $m_{3/2}=10$ GeV and
scales with gravitino mass as $\propto m_{3/2}^{-1}$.}
        \label{pot2}
    \end{center}
\end{figure}

It is also possible to consider limits arising in  Gauge-Mediated-Susy-Breaking (GMSB) models \cite{kolda}. Those alternative
scenarios, incorporating a natural suppression of the rate of
flavor-changing neutral currents due to the low energy scale, predict
the gravitino to be the Lightest Supersymmetric Particle (LSP). The
LSP is stable if R-parity is conserved. In this case, a limit was
obtained \cite{KBgrain} by requiring $\Omega_{3/2,0}<\Omega_{M,0}$,
{\it i.e.} by requiring that the current gravitino density  does not
exceed the matter density. It can easily be derived from the
previous method, by taking into account the dilution of gravitinos
in the period of PBH evaporation and the conservation of gravitino to keep
specific entropy ratio, that \cite{KBgrain}:
\begin{equation}\label{BetDM}\beta \leq \frac{\Omega_{M,0}}{N_{3/2}\frac{m_{3/2}}{M}\left(
\frac{t_{eq}}{t_{f}} \right) ^{\frac{1}{2}}}\end{equation} where
$N_{3/2}$ is the total number of gravitinos emitted by a PBH with
initial mass $M$, $t_{eq}$ is the end of the RD stage and
$t_f=max(t_{form},t_{end})$ is teh time at which a non-trivial equation of state for
the period of PBH formation is considered, {\it e.g.} a dust-like
phase which ends at $t_{end}$ \cite{polnarev1}. The limit
(\ref{BetDM}) does not imply the thermal equilibrium of the relativistic
plasma in the period before PBH evaporation and is valid even for
low reheating temperatures provided that the equation of state during
the preheating stage is close to relativistic. With the present
matter density $\Omega_{M,0}\approx 0.30$ \cite{wmap} this leads to
the limit shown in Fig. \ref{pot2} for $m_{3/2}=10$~GeV. Following
(\ref{BetDM}) this limit scales with gravitino mass as $
m_{3/2}^{-1}$. Models of gravitino dark matter with $\Omega_{3/2,0}
= \Omega_{CDM,0}$, corresponding to the case of equality in the
above formula, were considered in
\cite{Jedamzik1,Jedamzik11}.

\section {Massive Primordial Black Holes from the collapse of closed walls}\label{MBHwalls}

A wide class of particle models possesses a symmetry breaking
pattern, which can be effectively described by
pseudo-Nambu--Goldstone (PNG) fields and which corresponds to the
formation of an unstable topological defect structure in the early
Universe (see \cite{book2} for a review and references). The
Nambu--Goldstone nature of such an effective description reflects
the spontaneous breaking of a global U(1) symmetry, resulting in a
continuous degeneracy of vacua. The explicit symmetry breaking at a
smaller energy scale changes this continuous degeneracy to a discrete
vacuum degeneracy. The characteristics of the formed structures is  different
for phase transitions taking place during post-inflationary or
inflationary stages.
\subsection{Structures from a succession of U(1) phase transitions}\label{structures}

At high temperatures such a symmetry breaking pattern implies a
succession of second order phase transitions. In the first
transition, continuous degeneracy of vacua leads, at scales
exceeding the correlation length, to the formation of topological
defects in the form of a string network; in the second phase
transition, continuous transitions in space between degenerate
vacua form surfaces: domain walls surrounded by strings. This last
structure is unstable, but, as was shown in the example of the
invisible axion \cite{Sakharov2,kss,kss2}, it is reflected in the
large-scale inhomogeneity of the distribution of energy density of
coherent PNG (axion) field oscillations. This energy density is
proportional to the initial value of the phase, which acquires the dynamical
meaning of the amplitude of the axion field, when the axion mass is switched on
in the result of the second phase transition.

The value of the phase changes by $2 \pi$ around the string. This strong
nonhomogeneity of phase leads to a corresponding nonhomogeneity of the
energy density of coherent PNG (axion) field oscillations. The usual
argument (see e.g. \cite{kim} and references therein) is essential
only on scales corresponding to the mean distance between strings.
This distance is small, being of the order of the scale of the
cosmological horizon in the period during which PNG field oscillations
start. However, since the nonhomogeneity of the phase follows the
pattern of the axion string network, this argument misses large-scale
correlations in the distribution of the energy density of oscillations'.

Indeed, a numerical analysis of string networks (see review in
\cite{vs}) indicates that large string loops are strongly suppressed
and that a fraction of about 80\% of string length, corresponding to
long loops, remains virtually the same at all large scales. This
property is the other side of the well known scale invariant
character of string networks. Therefore the correlations in the energy
density should persist on large scales, as was shown in
\cite{Sakharov2,kss,kss2}.

The large-scale correlations in topological defects and their
imprints in primordial inhomogeneities constitute an indirect effect of
inflation, if phase transitions take place after the reheating of the
Universe. Inflation provides in this case identical conditions for the phase
transitions which takes place in causally disconnected regions.

If phase transitions take place during the inflationary stage, new forms of
primordial large scale correlations appear. The value of the phase after
the first phase transition is inflated over the region corresponding
to the period of inflation, while during
inflation fluctuations of this phase
 change  its initial value within regions of
smaller size. Owing to such fluctuations, for a fixed value of
$\theta_{60}$ in the period of inflation with {\it e-folding} $N=60$
corresponding to the part of the Universe within the modern
cosmological horizon, strong deviations from this value appear at
smaller scales, corresponding to later periods of inflation with $N
< 60$. If $\theta_{60} < \pi$, the fluctuations can move the value
of $\theta_{N}$ to $\theta_{N} > \pi$ in some regions of the
Universe. After reheating as a result of the second phase
transition these regions, which correspond to a vacuum with $\theta_{vac} =
2\pi$, are surrounded by the bulk of the volume with vacuum
$\theta_{vac} = 0$. As a result, massive walls are
 formed at the border between the
two vacua. Since regions with $\theta_{vac} = 2\pi$ are confined,
the domain walls are closed. As they fully enter the horizon,
closed walls can collapse into black holes.

  This mechanism can lead to the formation
of primordial black holes of arbitrarily large mass (up to the mass
of AGNs \cite{AGN}, see for latest review \cite{DER}). Such black
holes appear in the form of primordial black hole clusters,
exhibiting a fractal distribution in space
\cite{book2,KRS,Khlopov:2004sc}. Theyt can shed new light on the
problem of galaxy formation \cite{book2,DER1}.

\subsection {Formation of closed walls in an inflationary Universe}\label{walls}

To describe a mechanism for the appearance of massive walls of a
size essentially greater than the horizon at the end of inflation,
let us consider a complex scalar field with the
potential \cite{book2,AGN,KRS,Khlopov:2004sc}
\begin{equation}\label{V1} V(\varphi ) = \lambda (\left| \varphi
\right|^2  - f^2 /2)^2+\delta V(\theta ), \end{equation} where
$\varphi  = re^{i\theta } $. This field coexists with an inflaton
field which drives the Hubble constant $H$ during the inflational
stage. The term
\begin{equation} \label{L1} \delta V(\theta ) = \Lambda ^4 \left(
{1 - \cos \theta } \right), \end{equation} reflecting the
contribution of instanton effects to the Lagrangian renormalization
(see for example \cite{adams}), is negligible during the inflationary
stage and during some period in the FLRW expansion. The omitted term
(\ref{L1}) becomes significant when the temperature falls below the
values $T \sim \Lambda$. The mass of the radial field component $r$ is
assumed to be sufficiently large with respect to $H$, which means
that the complex field is in the ground state even before the end of
inflation. Since the term (\ref{L1}) is negligible during inflation,
the field has the form $\varphi \approx f/\sqrt 2 \cdot e^{i\theta }
$ and the quantity $f\theta$ acquires the meaning of a massless field.


At the same time,  the well established behavior of quantum field
fluctuations on de Sitter background \cite{Star80} implies that
the wavelength of a vacuum fluctuation of every scalar field grows
exponentially, having a fixed amplitude. Namely, when the wavelength
of a particular fluctuation, in the inflating Universe, becomes
greater than $H^{-1}$, the average amplitude of this fluctuation
freezes out at some  non-zero value because of the large friction
term in the equation of motion  of the scalar field, whereas its
wavelength grows exponentially. Such a frozen fluctuation is
equivalent to the appearance of a classical field that does not
vanish after averaging over macroscopic space intervals. Because the
vacuum must contain fluctuations at every wavelength, inflation
leads to the  creation of more and more new regions containing a
classical field of specific amplitude with a scale greater than
$H^{-1}$. In the case of an effectively massless Nambu--Goldstone
field considered here, the averaged amplitude of the phase fluctuations
generated during each e-fold (time interval $H^{-1}$)  is given by
\beq \label{fluctphase} \delta \theta = H/2\pi f. \eeq Let us assume
that the part of the Universe observed inside the contemporary
horizon $H_0^{-1}=3000h^{-1}$Mpc was inflating, over $N_U \simeq 60$
e-folds, from a single causally connected domain of size $H^{-1}$, for
which the average value of the phase is $\theta_0$. When
inflation begins in this region, after one e-fold, the volume of the
Universe increases by a factor $e^3$ . The typical wavelength of the
fluctuation $\delta\theta$ generated during every e-fold is equal to
$H^{-1}$. Thus, the whole domain  $H^{-1}$, containing $\theta_{0}$,
after the first e-fold effectively becomes divided into  $e^3$
separate, causally disconnected domains of size $H^{-1}$. Each
domain corresponds to an almost homogeneous  phase value
$\theta_{0}\pm\delta\theta$. Thereby, more and more domains appear
with time, in which the phase differs significantly from the initial
value $\theta_0$. A crucial point is the appearance of
domains with a phase $\theta >\pi$. Appearing only after a certain
period of time during which the Universe exhibited exponential
expansion, these domains turn out to be surrounded by a space with
phase $\theta <\pi$. The coexistence of domains with phases $\theta
<\pi$ and $\theta
>\pi$ leads, in the following, to the formation of
a large-scale structure of topological defects.

The potential (\ref{V1}) possesses a $U(1)$ symmetry which is
spontaneously broken at least after some period of inflation. Note
that the phase fluctuations during the first e-folds may, generally
speaking, transform into fluctuations of the cosmic
microwave radiation, and this will lead to restrictions on the
scaling parameter $f$. This difficulty can be avoided by taking into
account the interaction of the field $\varphi$ with the inflaton
field (i.e. by making parameter $f$ a variable~\cite{book2}). This
spontaneous breakdown is holding through the condition of smallness of
the radial mass, $m_r=\sqrt{\lambda_{\phi}}>H$. At the same time the
condition \beq\label{angularmass} m_{\theta}=\frac{2f}{\Lambda}^2\ll
H \eeq on the angular mass provides the freezing out  of the phase
distribution until some moment of the FRW epoch.  After the
violation of condition (\ref{angularmass}) the term (\ref{L1})
contributes significantly to the potential (\ref{V1}) and explicitly
breaks the continuous symmetry along the angular direction. Thus, the
potential (\ref{V1}) eventually has a number of discrete degenerate
minima in the angular direction at the points $\theta_{min}=0,\ \pm
2\pi ,\ \pm 4\pi,\ ...$ .

As soon as the angular mass $m_{\theta}$ is of the order of the
Hubble rate, the phase starts oscillating about the potential
minimum, with different initial values being in various space domains.
Moreover, in the domains with the initial phase $\pi <\theta < 2\pi
$, the oscillations proceed around the potential minimum at $\theta
_{min}=2\pi$, whereas the phase in the surrounding space tends to a
minimum at the point $\theta _{min}=0$. At the end of the decaying
phase oscillations, the system contains domains characterized by the
phase $\theta _{min}=2\pi$ surrounded by space with $\theta
_{min}=0$. Apparently, if we move in any direction from the inside to the
outside of the domain, we will unavoidably pass through a point
where $\theta =\pi$ because the phase varies continuously. This
implies that a closed surface characterized by the phase $\theta
_{wall}=\pi$ must exist. The size of this surface depends on the
moment of the domain formation in the inflation period, while the shape
of the surface may be arbitrary. The key point for the
subsequent considerations is that the surface is closed. After
reheating of the Universe, the evolution of domains with the phase
$\theta >\pi $ proceeds on the background of the Friedman expansion
and is described by the relativistic equation of state. When the
temperature falls down to $T_* \sim \Lambda$, an equilibrium state
between the "vacuum" phase $\theta_{vac}=2\pi$ inside the domain and
the $\theta_{vac} =0$ phase outside of it is established. Since the
equation of motion corresponding to potential (\ref{L1}) admits a
kink-like solution (see \cite{vs} and references therein), which
interpolates between two adjacent vacua $\theta_{vac} =0$  and
$\theta_{vac} =2\pi$,  a closed wall corresponding to the transition
region at $\theta =\pi$ is formed. The surface energy density of a
wall of width $\sim 1/m\sim  f/\Lambda^2$ is of the order \footnote{The existence of such domain walls in the theory
of the invisible axion was first pointed out in
\cite{sikivieinvisible}.} of $\sim
f\Lambda ^2$ .

Note that if the coherent phase oscillations do not decay for a long
time, their energy density can play the role of CDM. This is the
case, for example, in the cosmology of the invisible axion (see
\cite{kim} and references therein).

It is clear that immediately after the end of inflation, the size of
domains which contain a phase $\theta_{vac} >2\pi$ 
exceeds by far the horizon size.  This situation is replicated in the size
distribution of vacuum walls, which appear at the temperature $T_*
\sim \Lambda$ whence the angular mass $m_{\theta}$ starts to build
up. Those walls, which are larger than the cosmological horizon,
still follow the general FLRW expansion until the moment when they
get causally connected as a whole; this happens as soon as the size
of a wall becomes equal to the horizon size $R_h$. Evidently,
internal stresses developed in the wall after crossing  the horizon
initiate processes tending to minimize the wall  surface. This
implies that the wall tends, first, to acquire a  spherical shape
and, second, to contract toward its centre. For simplicity, we will
consider below the motion of closed spherical walls~\footnote{The
motion of closed vacuum walls has been derived analytically in
\cite{tkachev,sikivie}.}.

The wall energy is proportional to its area at the instant of
crossing the horizon. At the moment of maximum contraction, this
energy is almost completely converted into kinetic energy
\cite{Rubinwall}. Should the wall at the same moment be localized
within the gravitational radius, a PBH is formed.

A detailed study of black hole (BH) formation was made in \cite{AGN}.
The results of these calculations are sensitive to the
parameter $\Lambda$ and to the initial phase $\theta _U$. As the
$\Lambda$ value decreases to $\sim 1$ GeV, still greater PBHs
appear with masses of up to $\sim 10^{40}$ g. A change in the
initial phase leads to sharp variations in the total number of black
holes. As was shown above, each domain generates a family of
subdomains in its close vicinity. The total mass of such a cluster
is only 1.5--2 times that of the largest initial black hole in this
space region. Thus, the calculations confirm the possibility of
formation of clusters of massive PBHs ( $\sim 100M_{\odot}$ and
above) in the earliest stages of the evolution of the Universe at a
temperature of $\sim 1-10$ GeV. These clusters represent stable
energy density fluctuations around which baryonic matter (and
cold dark matter) may concentrate in the subsequent stages,
followed by the evolution into galaxies.

It should be noted that additional energy density is supplied by
closed walls of small sizes. Indeed, because of the smallness of their
gravitational radius, they do not collapse into BHs. After several
oscillations such walls disappear, leaving coherent fluctuations of
the PNG field. These fluctuations contribute to a local energy
density excess, thus facilitating the formation of galaxies.

The mass range of formed PBHs is constrained by the fundamental
parameters of the model $f$ and $\Lambda$. The maximal BH mass is
determined by the condition that the wall does not dominate locally
before it enters the cosmological horizon. Otherwise, local wall
dominance leads to a superluminal $a \propto t^2$ expansion for the
corresponding region, separating it from the other parts of the
Universe. This condition corresponds to the mass \cite{book2}\beq
\label{Mmax} M_{max} =
\frac{m_{pl}}{f}m_{pl}(\frac{m_{pl}}{\Lambda})^2.\eeq The minimal
mass follows from the condition that the gravitational radius of BH
exceeds the width of the wall and is equal to \cite{KRS,book2}\beq
\label{Mmin} M_{min} = f(\frac{m_{pl}}{\Lambda})^2.\eeq

Closed wall collapse leads to a primordial gravitational wave spectrum, 
peaked at \cite{PBHrev} \beq
\label{nupeak}\nu_0=3\cdot 10^{11}(\Lambda/f){\rm Hz} \eeq with
energy density up to \beq \label{OmGW}\Omega_{GW} \approx
10^{-4}(f/m_{pl}).\eeq At $f \sim 10^{14}$GeV this primordial
gravitational wave background can reach $\Omega_{GW}\approx
10^{-9}.$ For the physically reasonable values of \beq
1<\Lambda<10^8{\rm GeV}\eeq the maximum of the spectrum is at
\beq 3\cdot 10^{-3}<\nu_0<3\cdot 10^{5}{\rm Hz}.\eeq Another
profound signature of the considered scenario consists in gravitational wave
signals from the merging of PBHs in the PBH cluster. These effects can
provide tests of the considered approach in the eLISA experiment.

\section{Antimatter in a baryon-asymmetric Universe?}\label{antimatter}

Primordial strong inhomogeneities can also appear in the baryon
charge distribution. The appearance of antibaryon domains in the
baryon asymmetrical Universe, reflecting the inhomogeneity of
baryosynthesis, is a profound signature of such a strong
inhomogeneity \cite{CSKZ}. In the model of
spontaneous baryosynthesis (see Ref. \cite{Dolgov} for a review) the
possibility of the existence of antimatter domains, surviving to the
present day in an inflationary Universe with inhomogeneous
baryosynthesis was considered in \cite{zil}.

The mechanism of
spontaneous baryogenesis \cite{dolgmain} implies the existence of a
complex scalar field $\chi =(f/\sqrt{2})\exp{(\theta )}$ carrying
the baryonic charge. The $U(1)$ symmetry, which corresponds to the
baryon charge, is spontaneously and explicitly broken. The explicit
breakdown of the $U(1)$ symmetry is caused by the phase-dependent term
 \beq\label{expl} V(\theta )=\Lambda^4(1-\cos\theta ).
 \eeq
The possible baryon- and lepton-number violating interaction
of the field $\chi$ with matter fields can have the following
structure \cite{Dolgov} \beq\label{leptnumb} {\cal
L}=g\chi\bar QL+{\rm h.c.}, \eeq where the fields $Q$ and $L$ represent
a heavy quark and a lepton, coupled to the ordinary matter fields.
\begin{figure}
\begin{center}
\includegraphics[scale=0.7]{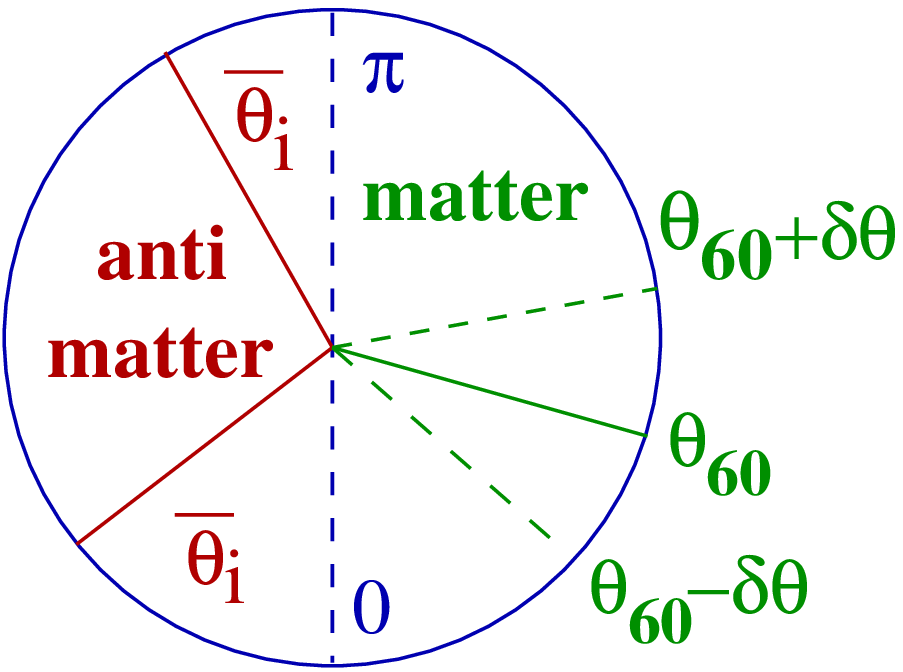}
\caption{ The inflationary evolution of the
phase (taken from \cite{ere}). The phase  $\theta_{60}$ sits in the range $[\pi ,0 ]$ at the
beginning of inflation and makes Brownian step
$\delta\theta_{eff}=H_{infl}/(2\pi f_{eff})$  at each e--fold. The
typical wavelength of the fluctuation $\delta\theta$ is equal to
$H^{-1}_{infl}$. The whole domain  $H^{-1}_{infl}$, containing the phase
$\theta_{N}$ gets  divided, after one e--fold, into  $e^3$ causally
disconnected domains of radius $H^{-1}_{infl}$. Each new domain
contains an almost homogeneous phase value
$\theta_{N-1}=\theta_{N}\pm\delta\theta_{eff}$. Every successive
e-fold, this process repeats in every domain.}
\label{pnginfl}
\end{center}
\end{figure}
In the early Universe, at a time when the friction term, induced by
the Hubble constant, becomes comparable to the angular mass
$m_{\theta}=\frac{\Lambda^2}{f}$, the phase $\theta$ starts to
oscillate around the minima of the potential and decays into
matter fields according to (\ref{leptnumb}). The coupling
(\ref{leptnumb}) gives rise to the following \cite{Dolgov}: as
the phase starts to roll down in the clockwise direction (Fig.~\ref{pnginfl}),  it preferentially creates an
excess of baryons over antibaryons, while the opposite is true as it
starts to roll down in the opposite direction.

The fate of such antimatter regions depends on their size. If their
physical size is larger than  the critical size
$L_c=8h^2$ kpc~\cite{zil}, they survive annihilation with surrounding matter.
The evolution of
sufficiently dense antimatter domains can lead to the formation of
antimatter globular clusters \cite{GC}. The existence of such a
cluster in the halo of our Galaxy should lead to the pollution of
the galactic halo by antiprotons. Their annihilation can reproduce
\cite{Golubkov} the observed galactic gamma background in the
range of tens to hundreds of MeV. The prediction of an antihelium component in
cosmic rays \cite{ANTIHE}, accessible to searches for
cosmic ray antinuclei in the AMS02 experiment, as well as
of antimatter meteorites \cite{ANTIME} provides a direct
experimental test for this hypothesis. The possibility of formation of dense antistars within an extension of the Affleck-Dine scenario of baryogenesis and the strategies for their search were recently considered in \cite{Blinnikov:2014nea}.

So primordial strong inhomogeneities in the distribution of
total, dark matter and baryon density in the Universe are a new
important phenomenon in cosmological models based on physics beyond the Standard model.

\section{Supersymmetry in the context of Cosmoparticle physics}\label{Discussion}

Observational cosmology offers strong evidence favoring the
existence of new physics, and suggests 
experimental approaches to their investigation.
Cosmoparticle physics \cite{ADS,MKH,book,newBook}, studying the
physical, astrophysical and cosmological impact of new laws of
Nature, explores new forms of matter and their physical
properties. The development of SUSY models follows the main principles of Cosmoparticle physics and offers a great challenge for
theoretical and experimental research. Physics of dark matter in all its aspects
plays important role in this process.

The necessity to extend the Standard model by supersymmetry has
serious theoretical reasons. Aesthetically, because it helps to achieve full
unification for the Standard model; practically,
because it removes its internal tensions. Supersymmetry can also provide a complete physical basis for cosmology. It can pretend to explain inflation, baryosynthesis and nonbaryonic dark
matter.  The white spots in the  representations  of supersymmetric models correspond to
new unknown particles. The extension of the symmetry of the gauge
group introduces new gauge fields, mediating new
interactions. Global symmetry breaking results in the existence of
Goldstone boson fields.

In particle physics, direct experimental probes for the predictions
of particle theory are the most attractive. The predicted  supersymmetric partners of known particles are accessible to experimental search
at accelerators if their masses are within
a few-TeV range. However, the predictions related to higher
energy scale need non-accelerator or indirect means for their
test.

Cosmoparticle physics offers complementary tools via indirect
and non-accelerator direct searches for new physics and its possible
properties. In experimental cosmoarcheology, data can be
obtained, which links the new physics
with astrophysical observations (see \cite{Cosmoarcheology}). In
experimental cosmoparticle physics the parameters, fixed from the
consitency of cosmological models and observations, define the
level at which the new types of particle processes should be
searched for (see \cite{expcpp}). These basic principles of cosmoparticle physics have been widely implemented in the development of supersymmetric models.

The theories of everything should provide a complete physical
basis for cosmology and naturally involve supersymmetry. The problem is that string theory
\cite{Green} is now in the form of "theoretical theory", for which many doubt that
experimental probes exist. cosmoparticle physics can remove these doubts.

For a long time, scenarios with Primordial Black holes belonged
dominantly to cosmological "fantasies", as they
provided restrictions on physics of the very early Universe from the
contradiction of their predictions with observational data. Even
this negative type of information makes PBHs an important
theoretical tool. Being formed in the very early Universe as an
initially nonrelativistic form of matter, PBHs should have increased
their contribution to the total density during the RD stage of
expansion, while the effect of PBH evaporation should have strongly
increased the sensitivity of astrophysical data to their presence.
Indeed, astrophysical constraints on hypothetical sources of cosmic or
gamma rays, 
on light element abundances and on the spectrum ofthe CMB can be linked to
restrictions on superheavy particles in the early Universe and on first- and second-order
phase transitions, thus making astrophysical data a sensitive probe of
particle symmetry structure and of the pattern of its breaking at superhigh
energy scales.

The gravitational mechanism of particle creation in PBH evaporation
makes evaporating PBHs a unique source of any species of particles
which can exist in our space-time. At least theoretically, PBHs can
be treated as the source of such particles, which are strongly
suppressed in any other astrophysical mechanism of particle
production, either due to a very large mass of these species, or
owing to their superweak interaction with ordinary matter.

By construction, astrophysical constraints exclude effects predicted
to be larger than observed. At the edge such constraints convert
into an alternative mechanism for the observed phenomenon. At some
fixed values of parameters, the PBH spectrum can play a positive role
and shed new light on old astrophysical problems.

Common sense dictates that PBHs should have small sub-stellar
mass. The formation of PBHs within the cosmological horizon, which was very
small in the very early Universe, seems to argue for this viewpoint.
However, phase transitions during the inflationary stage can provide spikes
in the spectrum of fluctuations at any scale, or lead to the formation of
closed massive domain walls of any size.

In the latter case, the existence of primordial clouds of massive black holes around an
intermediate-mass or supermassive black hole is possible. Such
clouds have a fractal spatial distribution. This
approach suggests a radically new scenario for galaxy
formation. Traditionally, the Big
Bang model assumes a homogeneous distribution of matter at all
scales, whereas the appearance of observed inhomogeneities is
related to the growth of small initial density perturbations.
However, the analysis of the cosmological consequences of the
particle theory indicates the possible existence of strongly
inhomogeneous primordial structures in the distribution of both
dark matter and baryons. These primordial structures represent a new
factor in galaxy formation theory. Topological defects such as the
cosmological walls and filaments, primordial black holes, archioles
in the models of axionic CDM, and inhomogeneous
baryosynthesis (leading to the formation of antimatter domains in
a baryon-asymmetric Universe
\cite{book,newBook,book2,CSKZ,zil,dolgmain,GC,
Golubkov,ANTIHE,ANTIME,sb,exl1,exl2,crg,kolb,we}) are an
incomplete list of possible primary inhomogeneities inferred from the
existing elementary particle models.

We can conclude that, within the modern cosmological paradigm, from 
the very beginning to the present time, the evolution of the
Universe was governed by physical laws which we still don't fully know.
These laws must come from a fundamental particle symmetry beyond the Standard model and they imply
the use of methods of cosmoparticle physics for their study.
Cosmoparticle physics originates from the well-established
relation between the microscopic and the macroscopic descriptions in
theoretical physics. This is reminiscent of the links between statistical physics
and thermodynamics, or between electrodynamics and the theory of the
electron. At the end of the XXth Century, a new instance of this kind of 
relationship was realized. It came both from the cosmological
necessity to go beyond the world of known elementary particles to settle the physical grounds for inflationary cosmology with
baryosynthesis and dark matter, and from the necessity for
particle theory to use cosmological tests as an important and in
many cases unique way to probe its predictions. The development of supersymmetric models
perfectly reflects this direction of fundamental knowledge.

\section*{Acknowledgments}
I express my gratitude to Prof. D.Cline for the kind invitation to contribute to this Special issue and to all my co-authors of the original papers, on which the present review is based. I am grateful to Prof. J.R. Cudell for reading the manuscript and valuable comments. The work on initial cosmological conditions was supported by the Ministry of Education and Science of Russian Federation, project 3.472.2014/K and the work on the forms of dark matter was supported by grant RFBR 14-22-03048.
\section*{Conflicts of Interest}

The author declares no conflict of interest.

\end{document}